 \documentclass[11pt,referee]{aa}
 \usepackage{graphics}
 \def\bmr{\hbox{\it B--R$\,\,$}}
 \def\deg{${}^\circ$}
 \def\Deg{${}^\circ$\llap{.}}
 \def\Min{${}^{\prime}$\llap{.}}
 \def\min{${}^{\prime}$}
 \def\Sec{${}^{\prime\prime}$\llap{.}}
 \def\hr{${}^{\rm h}$}
 \def\mn{${}^{\rm m}$}
 \def\sc{${}^{\rm s}$}

\begin{document}

     \thesaurus{(10             %
                10.15.2;       
                08.06.3;       
                08.12.3)}     

   \title{Photographic photometries and astrophysical parameters of the open clusters NGC1750 and NGC1758
        \thanks{This project is supported partly by the National Natural Science Foundation of
        China with Grant No. 19673012, 19603003 and 19733001, and in part by the astronomical
        foundation of Astronomical Committee of CAS.}}

      \author{ K.P.Tian\inst{1,2,3,} \and C.G.Shu\inst{1,2,3}
  \and J.L.Zhao\inst{4,1,2,3} \and P.B.Stetson\inst{5} \and C.Jordi\inst{6}
  \and D.Galad\`{i}-Enr\`{i}quez\inst{6}}
     \offprints{K.P. Tian, tkp@center.shao.ac.cn}

 \institute{ Shanghai Astronomical Observatory, CAS, Shanghai
 200030, and CAS-PKU Joint Beijing Astrophysical Center, P.R.China
 \and National Astronomical Observatories, CAS, P.R.China
 \and Joint Lab of Optical Astronomy, CAS, P.R.China
 \and CCAST(WORLD LABORATORY) P.O. Box 8730, Beijing, 100080, P.R.China
 \and Dominion Astrophysical Observatory, Herzberg Institute of Astrophysics, National Research
    Council of Canada, 5071 West Saanich Road, Victoria, British Columbia V8X~4M6, Canada
 \and Dept. d'Astronomia i Meteorologia, Uni. de Barcelona, Avda. Diagonal 647,
    E-08028 Barcelona, Spain}

  \date{Received ? , 1999; accepted ?}

  \titlerunning{NGC1750 and NGC1758}

  \maketitle

 \begin{abstract}

    {\it BV\/} photographic photometries of 789 stars in the region of the
 open clusters NGC1750 and NGC1758 are derived from a set of 8 photometric
 plates. According to the astronomical data including proper motions, positions
 and membership probabilities of individual stars, several astrophysical parameters
 for these two clusters are determined, such as HR diagrams, ages, distances,
 luminosity functions, masses and kinematics, etc. It is found that the  distances are
 ($525\pm 48$pc)  and ($794\pm 73$pc) with ages of $1.5\times10^8$yr and $6.3\times10^8$yr for
 NGC1750 and NGC1758 respectively. Both clusters show no significant mass segregation effects
 because of their relatively young ages. The analysis of the proper motions imply that
 their velocity distributions are isotropy.\

 \ Keywords{ open cluster: individual -- NGC1750, NGC1758 -- photometry --
 parameters}
 \end{abstract}

 \section{Introduction}

    Open clusters, as systems of stars having a common origin, provide a
 very powerful tool on studying stellar evolution history.  The homogeneity of
 photometric characteristics of stars in a cluster and their dynamics
 indicate that the cluster stars should have formed from one and the same
 primordial cloud within a relatively short time scale. Therefore, almost
 all member stars in a cluster should have roughly same age and chemical
 composition.  Furthermore, open clusters can be used to understand both the
 formation and kinematics of the Galactic disk due to their wide age
 and mass distributions.  For these reasons, open
 clusters constitute one of the most important  research fields in
 observational and theoretical astronomy.  Ages, distances, masses,
 luminosity functions and mass functions of open clusters are their basic researches.

   The region of the open cluster NGC1750 has been paid more and more
 attention mainly because it is a complex area including two open clusters (NGC1750 and
 NGC1758) and the Taurus dark clouds, which located in the anticenter direction of the
 Galaxy. Cuffey(1937) was the first one to study
 this area systematically, obtaining extensive photographic photometry of
 stars in blue and red photometric bands to a limiting magnitude about
 $R\sim 14\,$mag. However he did not clearly point out the existence of two open
 clusters in this region. He considered the whole area as NGC1746 with
 distance about 590pc and \bmr$\,\approx\,$0.30mag.

    More recently, Galad\'{i}-Enr\'{i}quez et al. (1998a, hereafter GJTR; 1998b; 1998c)
 did a series of studies of the open clusters NGC1750 and
 NGC1758, in which {\it UBVRI\/} CCD photometry of 3224 stars within the
 45\min$\,\times\,$45\min\ area was presented. At the same time, they combined different
 plates from several observatories to obtain proper motions for 45036 stars,
 and {\it BVR\/} photographic photometry of 39762 stars within
 2\Deg3$\,\times\,$2\Deg3 in this area  was completed down to
 $V\,\sim\,18.5\,$mag. Several physical parameters for NGC1750 and NGC1758
 were discussed, including their positions, sizes, density profiles, extinctions,
 distances, ages, luminosity functions and masses, etc.
 Tian et al.\ (1998, hereafter TZSS) obtained high-precision proper motions
 and membership probabilities for 540 stars within 1\Deg5$\,\times\,$1\Deg5
 in this area using 20 plates taken over a period of up to 68 years from
 Z\v{o}-S\'{e} 40cm Astrograph in Shanghai Astronomical Observatory, Chinese Academy of Sciences. In
 that work the two open clusters, NGC1750 and NGC1758, are successfully
 separated from each other. The core radii of the clusters
 have been estimated to be 17\Min20 and 2\Min25 respectively.

     In this paper, we will present the results of photographic photometry of 789
 stars to a limit of $V\sim16\,$ mag within a 35\min$\,\times\,$35\min\
 area round the center of NGC1750. Combining the results of proper motions and
 membership probabilities for individual stars, we will more deeply study some
 astrophysical parameters of the two open clusters, including their H-R diagrams,
 distances, ages, masses, mass functions and kinematics. The
 program is as follows. In sec. 2, we will introduce the data
 reduction and compare with previous work. Detailed discussion on
 the basic astrophysical researches are presented in Sec.3.
 Summary is arranged in the final section.

 \section{Photographic Photometry}

   \subsection{Observations, measurements and data reduction}
  The photographic photometries in the $B$ and $V$ bands were obtained from plates
 taken during 1992-1993 with the 1.56 meter reflecting telescope of Shanghai
 Astronomical Observatory, Chinese Academy of Sciences.  The plates cover an area of about
 35\min$\,\times\,$35\min, centered at $\alpha_{2000}$=5\hr3\mn30\sc,
 $\delta_{2000}$=23\deg44\min, which includes the clusters of NGC1750 and
 NGC1758.  The standard plate/filter combinations are IIaO+GG385 for the $B$
 band and 156-01+GG495 for the $V$ band. The size of individual plates is
 160mm$\times$160mm with a scale of 13\Sec25/mm, and the exposure time of
 individual plates is 30 minutes except for two plates, which have exposures
 of only 16 minutes. All plate materials are listed in Table 1, in
 which the first column denotes the plate number; the second column gives the
 epoch; filter and emulsion are shown respectively in Column
 3 and 4; Column 5 is the exposure time; and the last two columns are
 the number of standard stars adopted and the reduced residual for
 each plate respectively (see below).

 \begin{table*}
 \begin{flushleft}
 \normalsize \bf  Table~1. The plate materials for NGC1750 and NGC1758 \\ [0.05mm]

  \end{flushleft}
          \label{Tabwater1}
       \[
             \begin {array} {c l c c c c c }
              \hline
       \noalign{\smallskip}

  {\parbox[t]{13mm}{\centering  plate}} &
  {\parbox[t]{13mm}{\centering  epoch}} &
  {\parbox[t]{13mm}{\centering  filter}} &
  {\parbox[t]{16mm}{\centering emulsion}} &
  {\parbox[t]{13mm}{\centering  exp.time \\ \mbox{\scriptsize (min)}}} &
  {\parbox[t]{13mm}{\centering  N}}  &
  {\parbox[t]{13mm}{\centering  $\sigma$ }}\\
  \noalign{\smallskip}
  \hline

 {\rm cl93008} & 1993.1.18 & {\rm GG}495  & 156-01  &  30  & 34  &  0.027 \\
 {\rm cl93011} & 1993.11.19 & {\rm GG}495 & 156-01 &  30  &  31  &  0.029  \\
 {\rm cl93017} & 1993.1.14  & {\rm GG}495 & 156-01 & 16 & 29  &  0.031 \\
 {\rm cl93018} & 1993.11.14 & {\rm GG}495 & 156-01 & 16 & 23   &  0.037\\
 {\rm cl93012} & 1993.1.25 & {\rm GG}385  & {\rm IIaO} & 30  & 23  & 0.043 \\
 {\rm cl93005} & 1993.1.17 & {\rm GG}385  & {\rm IIaO} & 30  & 29   &  0.031 \\
 {\rm cl92001} & 1992.11.29& {\rm GG}385  & {\rm IIaO} & 30  & 34  &  0.027 \\
 {\rm cl92002} & 1992.11.29& {\rm GG}385  & {\rm IIaO} & 30  & 31  &  0.046\\
 \hline

 \end {array}
  \]
 \end {table*}
 The density measurements of the whole areas of individual plates were done on
 the Photometric Data Systems (PDS) model 1010 automatic measuring
 machine at the Dominion Astrophysical Observatory (DAO) of Canada.
 The reductions of $B$,$V$ magnitudes were carried out according to the
 method presented by Stetson (1979). Because the range of candidate
 stars on each plate covers more than 7 magnitudes, the images of
 bright stars are generally saturated  when the densities of faint
 stars can be measured. The fits of Gaussian point-spread-functions (PSF)
 to the unsaturated stars are good.  However, the difference between the
 real PSF and Gaussian distribution of saturated stars are obvious,
 especially for stars in the cluster center region. Fortunately, Stetson (1979)
 has already concerned these two kinds of situations. The photographic
 magnitude index $\mu$ can be defined by the following equation

 \begin {equation}
   \mu = constant - 5\,{\rm Log}(D),
 \end {equation}
 where $D$ is a measure of the total density of each star on a plate.

 The magnitude indices for each band are averaged
 together among all plates, and the individual plates are
 transformed to this average plate by the least-squares method. At
 the same time, these fittings permit the standard error of a magnitude index
 on a typical plate to be estimated from the transformation
 residuals.

   The Johnson-system {\it BV\/} photometric magnitudes are obtained
 by means of 141 secondary standard stars in this region with {\it BV\/} CCD
 photometry presented by GJTR before their paper was published. The magnitudes of these stars are all
 brighter than magnitude 14 in both the $B$ and $V$ bands. In addition, we
 choose  about 40 standard stars fainter than magnitude 14 from Table 5
 of GJTR, which have a homologous distribution on our plates. Our photographic  magnitude transformation equations are cubic
 polynomials in $B$ and $V$, with a linear color term in each band. Generally,
 the candidate stars of the photometric standards should obey two
 principles (see Shu et al. 1998 for details): (1) the
 photometric standards should be distributed as homogeneously as
 possible along the magnitude interval covered on the
 plates; (2) they should be well isolated. The number of
 the standard stars used in the transformation of each plate and
 the residual rms are listed in Column 6 and 7 of Table~1 respectively.

    The total number of stars we measured in the present study is 789. Their
 $B$ and $V$ magnitudes extend down to about 16.5$\,$mag with accuracies
 estimated by both  averaging the individual measurements of each star and
 considering the weighted internal error of the profile fitting.  Because the
 centers of the available photographic plates are not exactly the same, some
 stars with large fitting errors must be discarded.  At the same time, it is
 worthy noting that {\it BV\/} magnitudes of some stars are taken from only
 one or two plates. As a result, all 789 stars have $V$ magnitudes, with 540
 stars among them measured on three or four plates, and only 653 stars
 have $B$ magnitudes, among which 481 were measured on three or four plates. The
 final accuracy for the stars which are measurable on at least three
 plates is given as a function of apparent visual magnitude in Table~2 and
 illustrated in Figure~1. One can see that average accuracies are almost the same
 for different magnitudes except for the faint end, which show larger
 scatters.

 \begin{figure*}


   \includegraphics{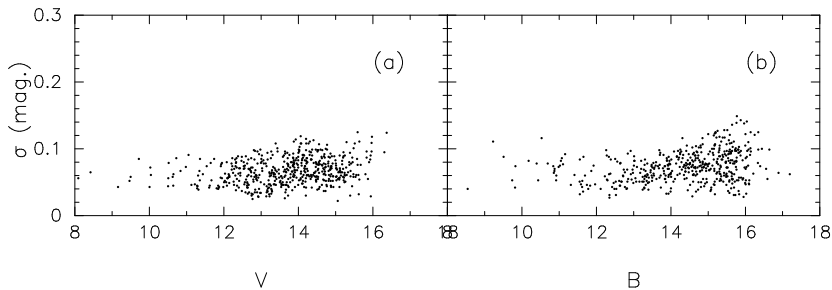}

    \caption{The accuracies of stars measured on at least
               3 plates as a function of apparent magnitude:
             (a) $V$ band  (b) $B$ band  }
          \label{FigSpectrum}
 \end{figure*}

 \begin{table}
 \normalsize \bf  Table 2. Number of stars observed (N) and\\
           averaged accuracy  $(\sigma)$ at
           different magnitude\\
            intervals\\ [0.05mm]
 \label{Tabwater2}
   \[
 \begin {array}  {   c   c   c  c  c  }
 \hline

    \rm magnitude  & \hspace*{17mm} \rm V &   & \hspace*{17mm} \rm B  &  \\

     \rm  range(mag)   & \rm   N   &  \sigma \rm(mag)  &  \rm N  &   \sigma \rm(mag)\\
 \hline
     \leq 10.  &   6  &  0.061   &   6  &  0.073   \\
  10.0 - 11.0    &  12  &  0.064   &  12  &  0.074  \\
  11.0 - 12.0    &  29  &  0.059   &  22  &  0.063  \\
  12.0 - 13.0    & 107  &  0.063   &  44  &  0.062  \\
  13.0 - 14.0    & 143  &  0.067   &  88  &  0.067 \\
  14.0 - 15.0    & 164  &  0.075   & 120  &  0.077 \\
  15.0 - 16.0    &  77  &  0.074   & 158  &  0.085    \\
  16.0 - 17.0    &   2  &  0.110   &  31  &  0.088 \\
  \hline
  \rm total          & 540  &  0.070   & 481   & 0.077   \\
  \hline

 \end {array}
  \]
 \end {table}

    The accuracies of the derived magnitudes depend on the film of
 the plates and its homogeneity, the magnitude range of the standard stars used in
 the reductions, and the uncertainty resulting from  transforming to the
 standard system. In general, it is difficult to obtain an accuracy of
 photographic photometry better than 0.1$\,$mag. In the present study, it
 must be pointed out that the average accuracies of V magnitudes of the 540
 stars and  of B magnitudes of the 481 stars, which are measured at least three as
 mentioned above, are $\pm0.070\,$mag and
 $\pm0.077\,$mag, respectively. The reasons are: (1) most of our standard
  stars have B and V CCD data with very high accuracies; (2)the standard
  stars we adopted have a homogeneous distribution in both  position and
  magnitude; (3)The method we chosen is reasonable.

 \begin{table}
 \normalsize \bf  Table 3. The photometry  Catalogue of the open clusters NGC1750\\
      and NGC1758, which is available in the electric form (see text)\\

 \scriptsize
 \label{Tabwater2}
   \[
 \begin {array}  { c c c c c c c c c c c c c }
 \hline\hline
 {\parbox[t]{8mm}{\centering No.}}&
 {\parbox[t]{19mm}{\centering R.A.(1950)\\ \mbox{$\mathrm{(^h\;\;\;^m\;\;\;^s)}$}}}&
 {\parbox[t]{18mm}{\centering DEC(1950) \\ \mbox{$(\degr\;\;\;\arcmin\;\;\;\arcsec)$}}}&
 {\parbox[t]{8mm}{\centering $\rm V$\\ }} &
 {\parbox[t]{8mm}{\centering $\rm \sigma_{V}$\\ }}&
 {\parbox[t]{8mm}{\centering $\rm N_V$ \\ }}&
 {\parbox[t]{8mm}{\centering $\rm B$ \\   }}&
 {\parbox[t]{8mm}{\centering $\rm \sigma_{B} $    }}&
 {\parbox[t]{8mm}{\centering $\rm N_B $\\   }}&
 {\parbox[t]{8mm}{\centering $\rm ID $ \mbox{\tiny (TZSS)}   }}&
 {\parbox[t]{8mm}{\centering $\rm P1 $\\   }}&
 {\parbox[t]{8mm}{\centering $\rm P2 $\\    }}&
 {\parbox[t]{8mm}{\centering $\rm ID $ \mbox{\tiny (GJTR)}   }}\\
      \noalign{\smallskip}

{\parbox[t]{10mm}{\centering \tt 1}}&
{\parbox[t]{20mm}{\centering \tt 2}}&
{\parbox[t]{17mm}{\centering \tt 3}}&
{\parbox[t]{15mm}{\centering  \tt 4}}&
{\parbox[t]{8mm}{\centering  \tt 5}}&
{\parbox[t]{11mm}{\centering  \tt 6}}&
{\parbox[t]{11mm}{\centering  \tt 7}}&
{\parbox[t]{11mm}{\centering  \tt 8}}&
{\parbox[t]{11mm}{\centering  \tt 9}}&
{\parbox[t]{11mm}{\centering  \tt 10}}&
{\parbox[t]{11mm}{\centering  \tt 11}}&
{\parbox[t]{11mm}{\centering  \tt 12}}&
{\parbox[t]{11mm}{\centering  \tt 13}}\\
     \noalign{\smallskip}
 \hline
 \small \rm

   $......$ & $......$ &      &     &     &    &    &    &    &    &    &    &        \\
  456 & 5\; 4\; 30.89 & +23\; 47\; 35.4 & 11.353 & 0.053 & 3 & 0.392 & 0.046 & 4 & 238 & 0.14 &0.84 & 2659  \\
  457 & 5\; 4\; 30.96 & +23\; 45\; 45.4 & 16.004 & 0.020 & 2 &       &       &   &    &      &     &  2660  \\
  458 & 5\; 4\; 31.02 & +23\; 50\; 34.1 & 14.410 & 0.020 & 2 & 0.834 & 0.065 & 4 &    &      &     &  2664 \\
  459 & 5\; 4\; 31.14 & +23\; 40\; 33.0 & 12.061 & 0.050 & 4 & 0.405 & 0.049 & 4 & 330&  0.96& 0.00&  2665 \\
  460 & 5\; 4\; 31.36 & +23\; 43\;  3.9 & 14.643 & 0.020 & 2 & 0.884 & 0.072 & 4 &    &      &     &  2671 \\
  461 & 5\; 4\; 31.37 & +23\; 49\; 32.6 & 15.256 & 0.020 & 2 & 1.038 &       & 2 &    &      &     &  2674 \\
  462 & 5\; 4\; 31.54 & +23\; 47\; 31.8 & 12.626 & 0.050 & 4 & 0.908 & 0.063 & 4 & 239&  0.00& 0.00&  2676 \\
  463 & 5\; 4\; 31.65 & +23\; 48\; 48.1 & 15.829 & 0.020 & 2 &       &       &   &    &      &     &  2680 \\
  464 & 5\; 4\; 31.78 & +24\;  1\; 24.5 & 12.444 & 0.060 & 4 & 1.037 & 0.074 & 4 &    &      &     &       \\
  465 & 5\; 4\; 31.88 & +23\; 45\; 36.2 & 12.929 & 0.028 & 4 & 0.523 & 0.065 & 4 & 261&  0.29& 0.70&  2688 \\
  466 & 5\; 4\; 31.95 & +23\; 33\; 18.9 & 15.385 & 0.094 & 3 & 0.726 &       & 2 &    &      &     &  2687 \\
  467 & 5\; 4\; 32.21 & +24\;  1\; 21.3 & 13.260 & 0.086 & 4 & 0.654 & 0.070 & 4 & 104&  0.06& 0.00&       \\
  468 & 5\; 4\; 32.39 & +23\; 43\; 25.3 & 15.179 & 0.020 & 0 & 1.022 &       & 2 &    &      &     &  2702 \\
  469 & 5\; 4\; 32.65 & +23\; 50\; 23.0 & 14.845 & 0.020 & 2 & 0.847 & 0.111 & 3 &    &      &     &  2707 \\
  470 & 5\; 4\; 32.79 & +23\; 36\; 51.3 & 13.227 & 0.029 & 4 & 0.710 & 0.033 & 4 & 372&  0.67& 0.00&  2708 \\
  471 & 5\; 4\; 32.94 & +23\; 50\;  8.5 & 13.864 & 0.086 & 4 & 0.789 & 0.073 & 4 & 198&  0.07& 0.92&       \\
  472 & 5\; 4\; 32.98 & +23\; 35\; 59.4 & 10.835 & 0.067 & 3 & 0.271 & 0.077 & 4 &  386&  0.96& 0.00&  2711 \\
  473 & 5\; 4\; 33.02 & +23\; 50\;  9.5 & 13.857 & 0.059 & 4 & 0.753 & 0.061 & 4 &     &      &     & 2716 \\
  474 & 5\; 4\; 33.11 & +23\; 38\; 27.1 & 15.071 & 0.065 & 3 & 0.840 &       & 2 &     &      &     & 2717 \\
  475 & 5\; 4\; 33.20 & +23\; 43\; 52.9 & 15.318 & 0.020 & 2 &       &       &   &     &      &     & 2721 \\
     ...... & ...... &      &     &     &    &    &    &    &    &    &    &        \\
 \hline

   \end{array}
      \]
    \end {table}
   \normalsize
     The final reduced photometric results for individual stars in the region
 of NGC1750 and NGC1758 are given in Table 3, which is available only in
 electronic form. We present a small part of Table 3 here as an example.
  Column 1 in Table 3 is the ordinal star number in
 order of increasing right ascension; Column 2 and column 3 present
 the equatorial coordinates of J2000; Column 4-6 and 7-9 are $V$ and
 $B$ magnitudes with corresponding standard error and
 the number of measured plates respectively. The 10th column lists the identification of TZSS.
 The next two columns, 11 and 12, show the membership probabilities of individual stars in
 NGC1750 and NGC1758 taken from TZSS. The cross-identification with GJTR
 (their ordinal star number) is given in the last column.

 \subsection{Comparisons}

     Here we estimate the external error of our $V$ and $B$ magnitudes
 through comparison with GJTR's CCD photometry. There are 448 and 347 common
 stars with $V$ and $B$ magnitudes respectively between our Table 3 and
 GJTR's photometry catalogue. Their differences $\Delta{V}$ and $\Delta{B}$
 in $V$ and $B$ bands as a function of magnitude
 for these stars are listed in Table 4 and shown in Figure 2. The mean
 differences in $V$ and $B$ magnitude are both
 $\pm 0.045\,$mag. It can be found clearly that this mean difference is small
 than the internal accuracy of stars available on at least three plates. It must be
 emphasized that this difference is not the external accuracy. It is because that
 our result is reduced from that of GJTR which has a very good accuracy. Furthermore,
 this also implies that the reduced method we adopted is reasonable. In fact, the
 mean difference of the 540 and 481 stars available on at least three plates in $B$
 and $V$ band respectively are larger than above we estimated in Table 2.
 It is consistent with the normal principle that the external accuracy must
 be worse than the internal accuracy.
 Thus the difference seems better

 \begin{figure*}


  \includegraphics{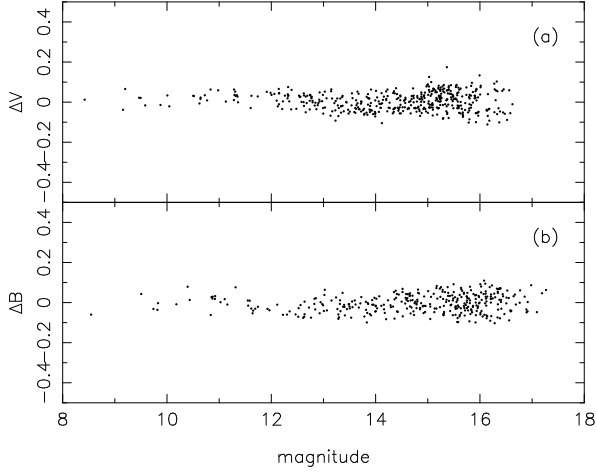}

    \caption{The magnitude differences between our and GJTR's
             results: (a) $V$ band  (b) $B$ band  }

          \label{FigSpectrum}
 \end{figure*}

 \begin{table}
 \normalsize \bf Table 4.  The magnitude differences between this work and GJTR's results\\  [0.05mm]
 \label{Tabwater4}
 \[
 \begin {array}  { c  c  c  c  c  c  }
 \hline

     \rm magnitude & \hspace*{17mm} \rm V &   & \hspace*{17mm} \rm B  &  \\

      \rm   range(mag)  &  \rm N   &  \Delta{V} \rm(mag)  & \rm  N  &   \Delta{B} \rm(mag) \\
 \hline

     \leq   10.0 &   7  &  0.032   &   5   &  0.039   \\
     10.0 - 11.0 &  11  &  0.035   &   9   &  0.039  \\
     11.0 - 12.0 &  14  &  0.037   &  14   &  0.033  \\
     12.0 - 13.0 &  53  &  0.033   &  22   &  0.045  \\
     13.0 - 14.0 &  73  &  0.040   &  45   &  0.041 \\
     14.0 - 15.0 & 100  &  0.037   &  68   &  0.037 \\
     15.0 - 16.0 & 160  &  0.053   & 108   &  0.046    \\
      \geq  16.0 &  30  &  0.064   &  76   &  0.055 \\
 \hline
       \rm  total   & 448  &  0.045   & 347   & 0.045   \\
  \hline
 \end {array}
 \]
 \end {table}

    In Figure 3 we compare the proper motions in the $x$ and $y$ directions
 for 517 stars common between our and GJTR's proper motion catalogs. The two
 astrometric results are obtained from different plate sets,
 reference stars of proper motions and reduce method.  It is found
 clearly that a fairly good linear relation exists in both components of the
 proper motions, but the slope is not unity. Although the proper motion in
 the present work is slightly larger than that of GJTR in both directions, which
is due to the different reference frames adopted, it can be
concluded that the two sets of results are consistent with each
other, and there will be no significant difference in the results
of membership determination because of the linear transformation.

 \begin{figure*}


   \includegraphics{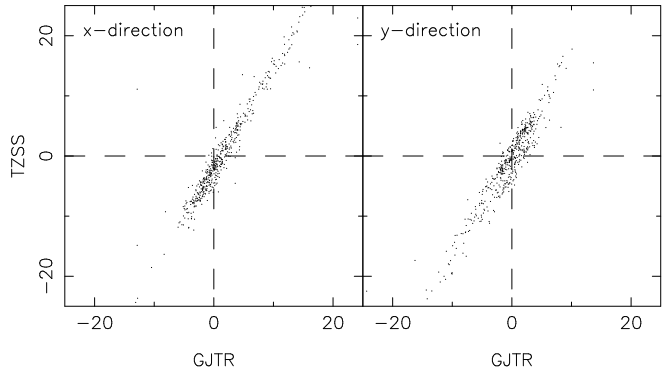}

    \caption{The comparions of the proper motions (in mas/yr)
               between this work and GJTR. }

          \label{FigSpectrum}
 \end{figure*}

 \section{Physical Parameters of NGC1750 and NGC1758}

    In order to investigate the basic astrophysical parameters of these two
 clusters, we must construct samples of members with positions, kinematics, membership
 probabilities and photometries  available for individual stars. In the present
 work, there are 504 stars with both $B$, $V$ photometry and proper motion data, for which
 positions, proper motions and membership probabilities can be obtained
 from our previous work (TZSS). $B$ and $V$ magnitudes for 238
 stars among them are taken from Table~3 and those for the remaining 266 are
 taken from the photometries done by GJTR. The sums of the membership
 probabilities for these 504 stars belong to NGC1750 and NGC1758 are 314 and 28
 respectively. Meanwhile, the numbers of stars with
 membership probabilities higher than 0.7 for NGC1750 and NGC1758 are 311 and 23. We reasonably
 choose these 311 and 23 stars as  our selected samples to analyze the CM diagram
 of the two clusters,  in order to obtain the distances,
 ages and the kinematics of the individual clusters. On the other hand, all 504 stars
 are used to study the luminosity functions and mass functions (see below).

   \subsection{Color-magnitude diagrams, distances and ages}

   The CM diagram offers a powerful diagnostic of the evolutionary state of
 an open cluster.  Because the locations of open clusters tend to be close
 to the plane of the Galaxy, their CM diagrams are liable to be heavily
 contaminated by unrelated field stars, and some caution must be taken into account to minimize
 this contamination by selecting stars on the basis of their kinematics, or
 by selecting only those stars with colors that are consistent with objects
 that have been reddened by the dust between us and the cluster.
 Figure 4 shows the observed CM  diagrams of NGC1750 and NGC1758,
 respectively, based on the sample described above. The dots
 denote stars with $B$ and $V$ taken from the present work and the open circles
 denote the stars with photometries from GJTR. It can be seen that both
 observational color-magnitude diagrams show fairly clear main sequences.
 Toward the bottom of the diagram, the main sequence becomes boarder
 for NGC1750, with a width too large to be attributed to
 observational errors, which reflects the containination of field stars

 \begin{figure*}


   \resizebox{\hsize}{!}{\includegraphics{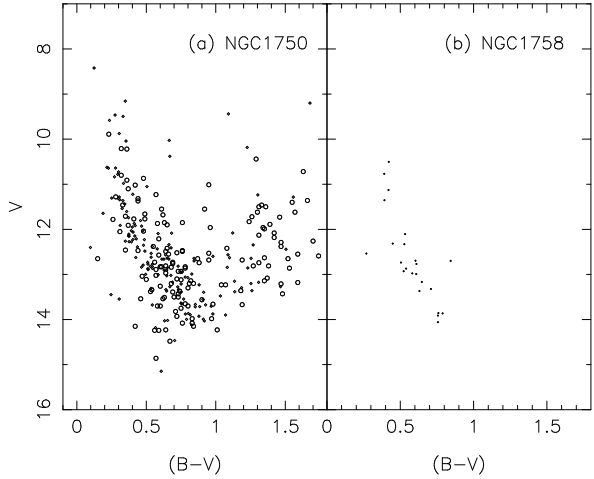}}

    \caption{The CM diagrams of NGC1750 and NGC1758 (see text).}

          \label{FigSpectrum}
 \end{figure*}

   In general, we do not know the cluster distances, so we cannot plot
 the CMDs on an absolute-magnitude scale.
 However, most cluster sizes are sufficiently small
 relative to their distance, so we can assume as usual that all
 stars belonging to a cluster lie at the same distance. To reduce
 contamination of field stars as much as possible, we trace the
 CM diagrams obtained by selecting stars with membership
 probabilities larger
 than 0.90, as shown in Fig.~5.  It can be found that Fig.5
  is much tighter than Fig.4. A careful discussion of
 the color excess due to dust absorption in front of these two clusters
 has been presented by GJRT. They found that the interstellar medium is
 relatively transparent toward the two clusters and the Johnson color
 excess for both of them is $\overline{E(B-V)}=(0.34\pm0.07)\,$mag,
 which corresponds to an extinction value of $\overline{A_{\rm
 v}}=(1.1\pm0.2)\,$ mag. Based on these observational properties
 and the empirical ZAMS (Mermilliod 1981; Schaller et al. 1992), we
 can derive the distance modular  of 8.60mag for NGC1750  and 9.50mag
 for NGC1758, which correspond to the distances of ($525\pm 48$pc) for NGC1750 and ($794\pm 73$pc)
 for NGC1758 with core radii of 2.6$\,$pc and 0.5$\,$pc respectively(TZSS).

 The age distribution of open clusters plays an important role in many
 astrophysical researches, which can be used to estimate the
 lower limit for the age of the Galactic disk (Grenon, 1989), investigate
 the formation and evolution, especially the star formation history
 of our disk, as well as its dynamics (Janes \& Phelps, 1994; Shu
 et al 1996). There are various methods to estimate ages of open
 clusters, which can lead to a significant scatter of the results
 for individual cluster.  This is because of the differences
 among isochrone fitting, conversion from theoretical to observed
 stellar parameters, and so on. The most popular method adopted up to today
 is the fitting of theoretical isochrones to the observed CM diagram.
 The age determination of NGC1750 and NGC1758 in the present study is
 relatively difficult because of the relatively small number of photographic
 plates and  some bright stars over-saturated, i.e., it is
 difficult to determine the turn-off points of these two clusters. Another
 reason is the relatively poor precision of the photographic photometries.
 Here, the same as GJTR did, we assume that the brightest stars on the main sequences of
 Fig.~5 denote the turn-off points of these two clusters, which are to be
 compared to the isochrones.  After comparing the observed color-magnitude diagrams with
 those of theoretical results given by Schaller et al. (1992) for solar metallicity, we get the
 estimated ages of $1.5\times10^8$yr for NGC1750 and $6.3\times10^8$yr for
 NGC1758, which are shown in Fig.~5

 \begin{figure*}


   \resizebox{\hsize}{!}{\includegraphics{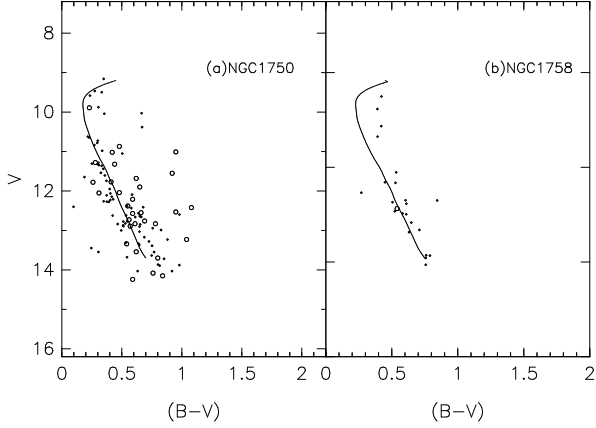}}

    \caption{The CM diagrams of NGC1750 and NGC1758, which are resulted from stars with
     membership probabilities $P\ge 0.90$, the solid lines denote ischrone fitting (see text).
     (a) NGC1750; (b) NGC1758}

          \label{FigSpectrum}
 \end{figure*}

   The lifetimes of main sequence stars as a function of their absolute visual
 magnitudes $M_V$ are
 also presented by Meynet et al (1993). The brightest star, for
 NGC1750, which is assumed to be on the MS,  is at $V=8.42$ mag, which corresponds
  to an age of about $4.1\times10^8
 \rm yr$. Similarly, the fact that the brightest star on the MS for NGC1758 has
 $V=10.77\,$mag leads to its age of $7.9\times10^8$\rm yr. According to the relation among stellar
 mass, lifetime and its $M_V$  given by Miller \& Scalo (1979), we can also obtain
 $1.6\times10^8$\rm yr and $7.8\times10^8$\rm yr for the ages of NGC1750 and
 NGC1758, respectively. If the relation between $M_V$ and lifetime is chosen as that presented by
 Mermilliod (1981), the ages of $3.6\times10^8\rm yr$ and $9.2\times10^8\rm yr$
 for NGC1750 and NGC1758 are inferred respectively. All adopted relations in present work are
 the average results. The main reasons for these different results are: (1) the different
  evolution tracks for stars resulted from different stars evolutionary model;
  (2) the different weight of metalliaities.  Combining all
 these results, we get the average age
 estimations for NGC1750  and NGC1758 should be $(2.2\pm1.0)\times10^8\,$yr and
 $(7.8\pm1.2)\times10^8\,$yr, respectively.

  \subsection{ Luminosity functions and mass functions}

  It is important to study luminosity functions (LFs) and mass
 functions (MFs) of individual open clusters because they can provide
 information about both the initial
 mass function (IMF) and cluster dynamical evolution. Conceptually, the
 simplest estimation of a cluster luminosity function is to count stars
 within the cluster. In order to reduce the contamination of field
 stars, the sum of stars' membership probabilities in different magnitude bins
 is one of the best to determine the
 luminosity functions $\Phi (\rm V)$ for individual clusters, i.e.,
  \begin {equation}
   \Phi(V)={\frac {\Sigma{P(i)}}{\Delta {V}}},
 \end {equation}

 \begin{figure*}

    \includegraphics{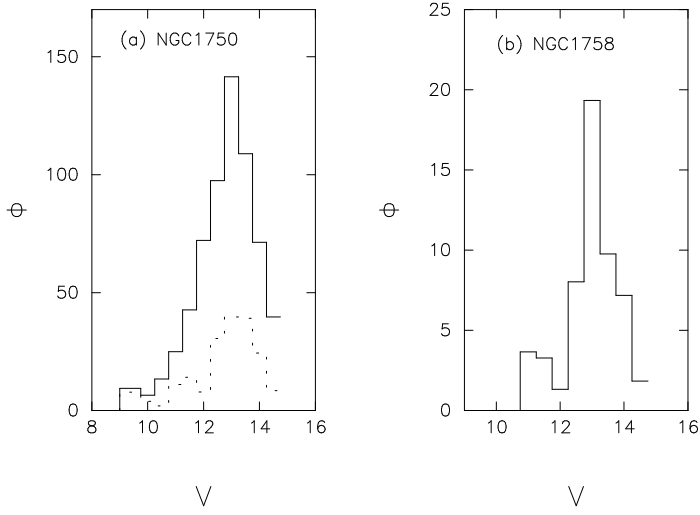}

    \caption{The observed LFs. (a) NGC1750, where the dots line denote the LF of
             the core region; (b) NGC1758.}

          \label{FigSpectrum} \end{figure*}
 where $P_c(i)$ is the
 membership probability of star $i$ within the magnitude range of V to
 $V+\Delta V$. Table~5 and Fig.~6 show the LFs for NGC1750 and NGC1758,
 respectively. One can see that there exists a peak for either clusters' LFs,
 which to some extent reflects the complete magnitudes of the samples.  The LF of the core region, which
 is within the center of  $2.6pc$, of NGC1750 is also given in Fig. 6 as a dotted line. We
 did no do the same thing for the NGC1758 because of its small number of
 member stars within the core. It is clear that  the profiles of luminosity functions in the central and
 whole observed region for NGC1750 are quite similar, i.e., there does not
 exist obvious mass segregation for NGC1750. The fact that the dynamical
 relaxation has not undergone thoroughly is consistent with its relatively young age (see last subsection).
 Combining the observed luminosity functions derived above and the
 mass-luminosity relations for main sequence stars given by Miller and
 Scalo (1979), we can infer the present-day mass functions of these two
 clusters, and the results are listed in Table~6 and also shown in Fig.~7
 respectively.  Here the average masses  in
 individual mass bins are weighted by membership probability, and $\Sigma
 P$ is summed over the stars in each bin as we did for their LFs, i.e.
 \begin {equation}
   \Psi(M/M_\odot)={\frac {\Sigma{P_i}}{\Delta {(M/M_\odot)_i}}},
 \end {equation}
 with $(M/M_\odot)_i={\frac {\Sigma{P_i (M_i/M_\odot)}}{\Sigma{P_i}}}$, here $M_i$ is the star mass with
 membership probabilities $P_i$ in the mass bin $\Delta {(M/M_\odot)}$.
 \begin{figure*}

   \includegraphics{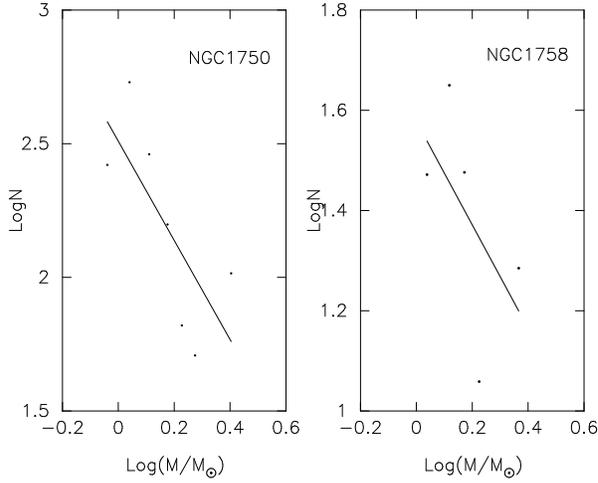}

    \caption{The observed present-day mass functions  for  NGC1750 and NGC1758 }

          \label{FigSpectrum}
 \end{figure*}

 \begin{table}
 \normalsize \bf  Table 5. The luminosity functions of\\
                 NGC1750 and NGC1758\\ [0.05mm]
 \label{Tabwater5}
  \[
 \begin {array} {c c || c c}
\hline
          \hspace*{2mm}\bf  NGC1750   &              & \hspace*{2mm} \bf  NGC1758          &        \\
                               &                 &                        &        \\

  \hspace*{4mm} \rm {V }  & \hspace*{12mm} \rm {\Sigma P }\hspace*{4mm} & \hspace*{5mm}\rm { M_V} & \hspace*{7mm} \rm {\Sigma }P\\
 \hline
    < 9.5 & \hspace*{11mm}  4.69 \hspace*{4mm}   & \hspace*{5mm}    < 11.0 \hspace*{5mm} & \hspace*{5mm}1.83  \\
   9.5-10.0 & \hspace*{10mm} 3.26  \hspace*{4mm}   & \hspace*{5mm}   11.0-11.5 \hspace*{5mm}& \hspace*{5mm} 1.62\\
  10.0-10.5 & \hspace*{10mm} 6.67  \hspace*{4mm}   & \hspace*{5mm}   11.5-12.0 \hspace*{5mm}& \hspace*{5mm} 0.66\\
  10.5-11.0 & \hspace*{10mm} 12.48 \hspace*{4mm}   & \hspace*{5mm}   12.0-12.5\hspace*{5mm}& \hspace*{5mm} 4.01 \\
  11.0-11.5 & \hspace*{10mm} 21.34 \hspace*{4mm}   & \hspace*{5mm}   12.5-13.0 \hspace*{5mm} &\hspace*{5mm}  9.67 \\
  11.5-12.0 & \hspace*{10mm} 36.07 \hspace*{4mm}   & \hspace*{5mm}   13.0-13.5 \hspace*{5mm} & \hspace*{5mm} 4.88 \\
  12.0-12.5 & \hspace*{10mm} 48.75 \hspace*{4mm}   & \hspace*{5mm}   13.5-14.0 \hspace*{5mm} & \hspace*{5mm} 3.59 \\
  12.5-13.0 & \hspace*{10mm} 70.76 \hspace*{4mm}   & \hspace*{5mm}     > 14.0  \hspace*{5mm}&  \hspace*{5mm}0.92 \\
  13.0-13.5 & \hspace*{10mm} 54.42 \hspace*{4mm}   &         &     \\
  13.5-14.0 & \hspace*{10mm} 35.65 \hspace*{4mm}   &         &     \\
  > 14.0   & \hspace*{10mm} 19.85 \hspace*{4mm}   &           &    \\
  \hline
  \end {array}
   \]
  \end {table}
  The slopes of the present-day mass functions of the two
 clusters are obtained by the least-squares linear regression.  The
 results are shown by means of Log-Log plots in Figure 7. The slopes
 are $(-1.85\pm0.19)$ and $(-1.18\pm0.33)$ with the correlation
 coefficients of 0.83 and 0.66 for NGC1750 and NGC1758
 respectively.Both clusters show the negative slopes. This also
 implies that they have not suffered the dynamical relaxation, which
 is consistent with the previous results.

 Furthermore, based on the M/L relation given by Miller \& Scalo
 (1979), the observed masses in the cluster region can be estimated
 to be about 390 $\rm M_{\odot}$ and 40$\rm M_{\odot}$ for NGC1750
 and NGC1758, respectively. Here, binary stars are not
 considered, so these results are probably underestimated.

 \begin{table}
 \normalsize \bf  Table 6. The mass functions of
                   NGC1750 and NGC1758\\ [0.05mm]
 \label{Tabwater6}
  \[
 \begin {array} {c c c ||c c c}
 \hline
                &     \bf  NGC1750        &           &              &  \bf NGC1758        &        \\
 \hline
 \rm    {bin}       &\rm {\overline {mass}}    &\rm {\Sigma P } & \rm {bin} & \rm {\overline {mass}} & \rm {\Sigma P}  \\
    \rm {(M/M_\odot)} & \rm {(M/ M_\odot)} &     & \rm {(M/M_\odot)} & \rm {(M/M_\odot)} &   \\
 \hline

        < 1.0      &  0.91      &  79.03  &   < 1.2       &  1.09      &   6.52 \\
      1.0 - 1.2    &  1.10      & 107.48  &  1.2 - 1.4    &  1.31      &   8.93\\
      1.2 - 1.4    &  1.29      &  57.82  &  1.4 - 1.6    &  1.49      &   5.99\\
      1.4 - 1.6    &  1.50      &  31.56  &  1.6 - 1.8    &  1.68      &   2.29\\
      1.6 - 1.8    &  1.69      &  13.21  &    > 1.8      &  2.32      &   3.47\\
      1.8 - 2.0    &  1.88      &  10.21  &               &            &     \\
          > 2.0    &  2.54      &  15.53  &               &            &    \\
  \hline
  \end {array}
   \]
  \end {table}
   \subsection{The kinematics }

 We might hope that direct studies of the kinematics of stars in NGC1750 and
 NGC1758 would reveal the effect of mass and space segregation. A
 reliable method for studying the kinematics of open clusters is based on
 proper motions of the member stars, which are  comparatively easy to be obtained.
 In our sample, the average accuracy of proper motions is
 $0.67mas\rm yr^{-1}$(TZSS), which corresponds to 1.7\rm {$\,$km$\,$s$^{-1}$} for
 the distance of NGC1750 and 2.5\rm {$\,$km$\,$s$^{-1}$} for the distance of
 NGC1758.  Considering the stars with membership probabilities greater than
 0.70, we estimate the intrinsic proper motion dispersions based on all the
 stars in the sample using the method outlined by Sagar \& Bhatt (1988). The
 dependences of the intrinsic velocity dispersions on stellar masses and
 distances from the cluster centers are listed in Table~7 and Table~8 for
 NGC1750 and NGC1758 respectively, where the radial distances of each star is
 measured from the centers of the two clusters determined by TZSS, and N denotes the star number we used.
 One can see in Table~7 that there is no statistically significant radial
dependence of the intrinsic proper motion dispersion $\sigma_{\mu}$. The values of $\sigma_{\mu}$ for
different radial shells are almost the same within their
uncertainties. It can also
 be seen from Table~7 that the intrinsic velocity dispersions $\sigma_\mu$ of
 different mass groups are almost the same.   This means that the present data provide litter evidences of mass
 segregations in these two young open clusters, which is consistent with the
 results we obtained above.  Even so, we can find from Table 7 that in the core region of NGC1750
 $(r<20 arcmin)$, the intrinsic velocity dispersions of the stars with larger  mass
 are smaller than the intrinsic velocities of the stars with smaller
 mass,  there exists some degree of both space and velocity
 mass segregation in the center region of NGC1750, where the dynamical relaxation is easy
 to undergo. But it is not clear for NGC1758 due to its small number of member stars.

 \begin{table}
 \normalsize \bf  Table 7. Dependence of intrinsic dispersion in proper motion
                       on stellar mass and radial distance for NGC1750\\ [0.05mm]
 \label{Tabwater7}
  \[
 \begin {array} { c c c c c }
 \hline

          \rm {Radius}   & \rm  { V}   &   \rm  {Mass}       &  \rm {\sigma _\mu}      &  \rm {N}\\

          \rm {(arcmin)} &   \rm  {(mag)}    & \rm {(\rm M/\rm M_{\odot})}  & \rm  {( "/100yr)} &   \\
 \hline

                        &   <  11.0    &  1.76 - 3.77     &        0.141\pm0.019     &  28 \\
                        & 11.0 - 12.0  &  1.34 - 1.76     &        0.175\pm0.016     &  59 \\
                        & 12.0 - 12.5  &  1.19 - 1.34     &        0.181\pm0.019     &  47 \\
                        & 12.5 - 13.0  &  1.05 - 1.19     &        0.179\pm0.015     &  68 \\
                        & 13.0 - 13.5  &  0.98 - 1.05     &        0.189\pm0.020     &  52 \\
                        &    > 13.5

                            &  0.73 - 1.05     &        0.132\pm0.012     &  57 \\
                        &              &                  &                          &     \\

               <10      &              &                  &        0.149\pm0.015     &  48 \\
              10 - 20   &              &                  &        0.166\pm0.015     &  62 \\
              20 - 30   &              &                  &        0.190\pm0.014     &  91 \\
              30 - 40   &              &                  &        0.189\pm0.016     &  70 \\
                >  40   &              &                  &        0.172\pm0.019     &  40 \\
                        &              &                  &                          &      \\

                < 10    &   < 12.5     &  1.19 - 3.77     &        0.109\pm0.015     &  25  \\
               10 - 20  &   < 12.5     &  1.19 - 3.77     &        0.138\pm0.019     &  25  \\
               20 - 30  &   < 12.5     &  1.19 - 3.77     &        0.189\pm0.021     &  39 \\
               30 - 40  &   < 12.5     &  1.19 - 3.77     &        0.197\pm0.025     &  30  \\
               > 40     &   < 12.5     &  1.19 - 3.77     &        0.158\pm0.029     &  15  \\
                        &              &                  &                          &      \\

               < 10     &   > 12.5     &  0.73 - 1.19     &        0.168\pm0.026     &  23 \\
               10 - 20  &   > 12.5     &  0.73 - 1.19     &        0.180\pm0.021     &  37 \\
               20 - 30  &   > 12.5     &  0.73 - 1.19     &        0.188\pm0.018     &  52 \\
               30 - 40  &   > 12.5     &  0.73 - 1.19     &        0.180\pm0.020     &  40 \\
                > 40    &   > 12.5     &  0.73 - 1.19     &        0.165\pm0.020     &  25 \\
  \hline
  \end {array}
   \]
  \end {table}

 \begin{table}
 \normalsize \bf  Table 8. Dependence of intrinsic dispersion in proper motion
                       on stellar mass and radial distance for NGC1758\\ [0.05mm]
 \label{Tabwater8}
  \[
 \begin {array} {c c c c c }
 \hline

       \rm { Radius}   &  \rm  {V}       & \rm  {Mass}       & \rm { \sigma _\mu}    &  \rm {N}\\

       \rm  { (arcmin) }& \rm {(mag)}    & \rm {(\rm M/\rm M_{\odot}) } & \rm {( "/100yr)} &   \\
 \hline

                        &              &                  &                          &     \\
                        &   < 12.8     &  1.41 - 2.61     &        0.123\pm0.025     & 12  \\
                        &   > 12.8     &  1.02 - 1.38     &        0.110\pm0.024     & 11  \\
                        &              &                  &                          &     \\
                < 2     &              &                  &        0.126\pm0.028     &  11  \\
                > 2     &              &                  &        0.113\pm0.023     &  12  \\
  \hline
  \end {array}
   \]
  \end {table}

     To gain information about the isotropy or anisotropy of the velocity
 distribution, the radial and tangential components $\sigma_{\mu
 r}$, $\sigma_{\mu t}$ of the intrinsic dispersions of proper
 motions as a function radius are calculated  and listed in Table
 9, where the units of radial distance $r$ and of proper motion
 dispersions $\sigma_{\mu}$ have been converted into pc and $\rm km
 s^{-1}$ respectively. The computation has been made for stars in the NGC1750
 and NGC1758 regions with membership probabilities higher than 0.7.  It can
 be found for both clusters that the ratios $\sigma_{\mu r}/\sigma_{\mu t}$ fluctuate around unity,
 which implies the absence of any significant evidence of velocity
 anisotropy. On the other hand, because the number of member stars is small, we can not get
 any certainly statistical results for NGC1758, but at least we can conclude
 that except the center region of NGC1750,
 there is no obvious velocity mass segregation or spatial mass segregation
 among the member stars of NGC1750, which suggests that this young open
 cluster has not reached energy equipartition.

 \begin{table}
 \normalsize \bf  Table 9. Dependence of radial and tangential intrinsic dispersion in proper motion
                       on stellar mass and radius\\ [0.05mm]
 \label{Tabwater}
  \[
 \begin {array} {c c c c c c }
 \hline

    \rm {Cluster}  & \rm {radius}    & \rm {\sigma_{\mu r}}   & \rm {\sigma_{\mu t}}  & \rm {\sigma_{\mu r}/\sigma_{\mu t}} &    N    \\
                  &  \rm {(pc)}             & \rm {(\rm kms^{-1})}   & \rm {(\rm kms^{-1})}     &                                &         \\
 \hline
    {\rm NGC1750} &        < 1.54         &    5.86\pm0.77     &    6.10\pm0.69    &    0.96\pm1.47 & 48  \\
                  &     1.54 - 3.07       &    5.22\pm0.55     &    4.08\pm0.48    &    1.28\pm1.72 & 62  \\
                  &     3.07 - 4.61       &    5.18\pm0.42     &    5.90\pm0.50    &    0.88\pm1.20 & 91  \\
                  &     4.61 - 6.15       &    5.40\pm0.55     &    5.22\pm0.55    &    1.03\pm1.20 & 70  \\
                  &      > 6.15           &    4.85\pm0.91     &    6.23\pm0.91    &    0.77\pm1.26 & 40  \\
                  &                       &                    &                   &                &    \\
    {\rm NGC1758} &      < 0.46           &    6.00\pm1.50     &    5.77\pm1.72    &    1.04\pm1.36 & 11   \\
                  &      > 0.46           &    1.74\pm0.97     &    3.16\pm1.93    &    0.55\pm0.94 & 12   \\
   \hline
  \end {array}
   \]
  \end {table}

 \section{Conclusion}

      In the present paper, based on the proper motions, photometries and membership
  probabilities of individual stars in the region of NGC1750 (TZSS, GJTR), we
  investigate basic astrophysical properties for two dynamically independent open
  clusters, NGC1750 and NGC1758. After detailed discussions on the photometric
  data and membership probabilities, the analysis samples with star number
  of 311 and 23 for NGC1750 and NGC1758 are constructed respectively. Comparing
  ZAMS (Mermilliod 1981; Schaller et al. 1992), we obtain the distances for
  these two clusters of $(525\pm 48)\,$pc for NGC1750 and $(794\pm 73)\,$pc
  for NGC1758 with their extinction being considered. Furthermore, many methods
  for the age determination are adopted to estimate the average ages of
  $(2.2\pm0.6)\times10^8 \rm yr$ and $(7.8\pm1.6)\times10^8 \rm yr$ with the
  observed masses 390$\,$M$_\odot$ and 40$\,$M$_\odot$ for NGC1750 and NGc1758 respectively.

     According to  the results of membership determination, luminosity functions and mass functions
  are given at the same time. It can be concluded that there exist no significant
  mass segregation effects for both clusters, which is consistent
  with  the fact that their dynamical relaxation have not undergone thoroughly.

     Finally, the velocity distributions of member stars for these two clusters
  are also discussed. It is found that both clusters seem to be isotropy
  in velocity space. Moreover, it is worth nothing that our statistical results could not
  be enough certain for NGC1758 because of its small number of member stars observed.

 \section*{Acknowledgements}
     The present work is part supported under the National Natural Science Fundation of China
 Grant No. 19673012 and 19733001 and by the astronomical fundation of Astronomical Committee of CAS.
 This work is also supported  in part under Joint Laboratory for Optional Astronomy of CAS.
 K.P.Tian and J.L.Zhao are grateful to  The National
 Research Council of Canada, which supported the living expenses  while
 they visited the Dominion Astrophysical Observatory.

\end{document}